\documentclass[a4paper,11pt,oneside]{article}

\usepackage[english]{babel}
\usepackage[T1]{fontenc}
\usepackage[utf8]{inputenc}

\usepackage[pdftex,unicode]{hyperref}
\hypersetup{pdftitle=The Pitfalls of Continuous Heavy-Tailed Distributions in High-Frequency Data Analysis}
\hypersetup{pdfauthor=Vladimír Holý}

\usepackage{tikz}

\usepackage{xcolor}
\definecolor{mycol}{rgb}{0.42,0.37,0.57}
\hypersetup{colorlinks=true, linkcolor=mycol, anchorcolor=mycol, citecolor=mycol, filecolor=mycol, urlcolor=mycol}

\usepackage[margin=60pt]{geometry}
\usepackage[authoryear]{natbib}

\usepackage{amsmath}
\usepackage{amssymb}
\usepackage{graphicx}
\usepackage[group-minimum-digits=3]{siunitx}
\usepackage{booktabs}
\usepackage{multirow}
\usepackage{tablefootnote}
\usepackage{float}
\usepackage{caption}

\usepackage[normalem]{ulem}

\usepackage{pifont}

\usepackage[super]{nth}

\begin{document}

\begin{center}
{\Large \bfseries The Pitfalls of Continuous Heavy-Tailed Distributions in High-Frequency Data Analysis}
\end{center}

\begin{center}
{\bfseries Vladimír Holý} \\
Prague University of Economics and Business \\
Winston Churchill Square 4, 130 67 Prague 3, Czechia \\
\href{mailto:vladimir.holy@vse.cz}{vladimir.holy@vse.cz} \\
\end{center}

\noindent
\textbf{Abstract:}
We address the challenges of modeling high-frequency integer price changes in financial markets using continuous distributions, particularly the Student's t-distribution. We demonstrate that traditional GARCH models, which rely on continuous distributions, are ill-suited for high-frequency data due to the discreteness of price changes. We propose a modification to the maximum likelihood estimation procedure that accounts for the discrete nature of observations while still using continuous distributions. Our approach involves modeling the log-likelihood in terms of intervals corresponding to the rounding of continuous price changes to the nearest integer. The findings highlight the importance of adjusting for discreteness in volatility analysis and provide a framework for incroporating any continuous distribution for modeling high-frequency prices.
\\

\\

\\

\section{Introduction}
\label{sec:intro}

Although originally developed to model daily volatility in financial markets, generalized autoregressive conditional heteroskedasticity (GARCH) models are also applied in high-frequency data analysis to capture intraday time-varying volatility (see, e.g., \citealp{Engle2002b}). Intraday volatility modeling dates back to \cite{Ghysels1998}, \cite{Engle2000}, and \cite{Meddahi2006}. Over the years, the intensity of trading has gradually increased, and the discreteness of prices has become more pronounced. To address this issue of discreteness, recent literature has proposed several dynamic models based on the Skellam distribution, such as \cite{Koopman2017a}, \cite{Koopman2018}, \cite{Alomani2018}, \cite{Goncalves2020}, \cite{Cui2021}, \cite{Doukhan2021}, \cite{Catania2022}, and \cite{Holy2024d}.

We take a different approach and address the issue of discreteness concerning the use of the continuous Student's t-distribution. In Section \ref{sec:data}, we outline an intraday analysis of stock price changes. For conciseness, we focus on the IBM stock; however, the empirical properties studied are widely observed across various stock markets. Empirical evidence for additional stocks is presented in Appendix \ref{app:further}. In Section \ref{sec:cont}, we demonstrate that the Student's t-distribution is unsuitable for analyzing integer price changes. The main reason is that its density function tends to degenerate into a shape of $\perp$, concentrated at a single point 0 with extremely heavy tails. In this sense, we build on the study of \cite{Holy2024d}, in which this behavior was observed but not examined in detail. As we further show, the main risk lies in the fact that results from dynamic models estimated using various R packages may initially appear reasonable but are based on incorrect estimates. In Section \ref{sec:mle}, we argue that continuous distributions can still be used, but they require a special method of estimation that accounts for the discrete nature of observations. We propose formulating the log-likelihood function in terms of the probabilities of continuous price changes falling into the interval corresponding to the rounding to the nearest integer that is actually observed. In Section \ref{sec:integer}, we estimate our GARCH-like model using the proposed interval maximum likelihood approach and find that it provides a comparable fit to the Skellam distribution. We conclude the paper in Section \ref{sec:con}.

The first contribution of this paper is the warning to practitioners not to rely on standard GARCH modeling with continuous distributions, especially those with heavy tails, as the results are uninformative at best and misleading at worst. The second contribution is the proposal of the simple modification to the maximum likelihood estimation procedure that accounts for the discreteness of data while still allowing the use of continuous distributions. The findings of our paper are relevant for researchers, risk managers, and traders who rely on these models for volatility forecasting, portfolio optimization, and derivative pricing.

\section{Data and Modeling Strategy}
\label{sec:data}

In this study, we analyze the IBM stock traded on the New York Stock Exchange (NYSE). The IBM stock was selected because many relevant studies feature it, including \cite{Engle2000}, \cite{Liu2012}, \cite{Catania2022}, and \cite{Holy2024d}. However, we stress that our core results are quite general, as the study revolves around two features that are widely observed across various financial markets---namely, the discreteness of price changes and the fact that zero is the most likely change. In Appendix \ref{app:further}, we also report results for three additional stocks---the MCD stock traded on the NYSE, and the CSCO and MSFT stocks traded on the NASDAQ. These results are generally consistent with those for the IBM stock.

Our data sample for the IBM stock consists of tick-by-tick transactions from all 252 trading days in 2024. The data source is Refinitiv Eikon. We perform standard data cleaning steps of \cite{Barndorff-Nielsen2009}. Specifically, we remove observations outside the standard trading hours of 9:30--16:00 EST, remove observations without recorded prices, and remove outliers with prices exceeding 10 mean absolute deviations within a rolling window of 201 observations. After data cleaning, we are left with over 15 million observations.

The tick-by-tick data have irregularly spaced observations, which pose some challenges. \cite{Engle2000} addressed this issue by using returns divided by the square root of trade durations, thus modeling the variance per time unit. As pointed out by \cite{Holy2024d}, this approach faces a significant issue for recent datasets due to the large number of zero or close-to-zero durations. Our data are recorded with millisecond precision, and 45 percent of transactions have zero durations. Although \cite{Holy2024d} proposed a model to address this issue, we decide to analyze the data at fixed frequencies for simplicity. We aggregate observations at 0.1, 1, 10, 60, and 300-second frequencies using the last tick method. We analyze price changes, i.e., the difference in successive prices in cents.

\begin{figure}
\centering
\includegraphics[width=\textwidth]{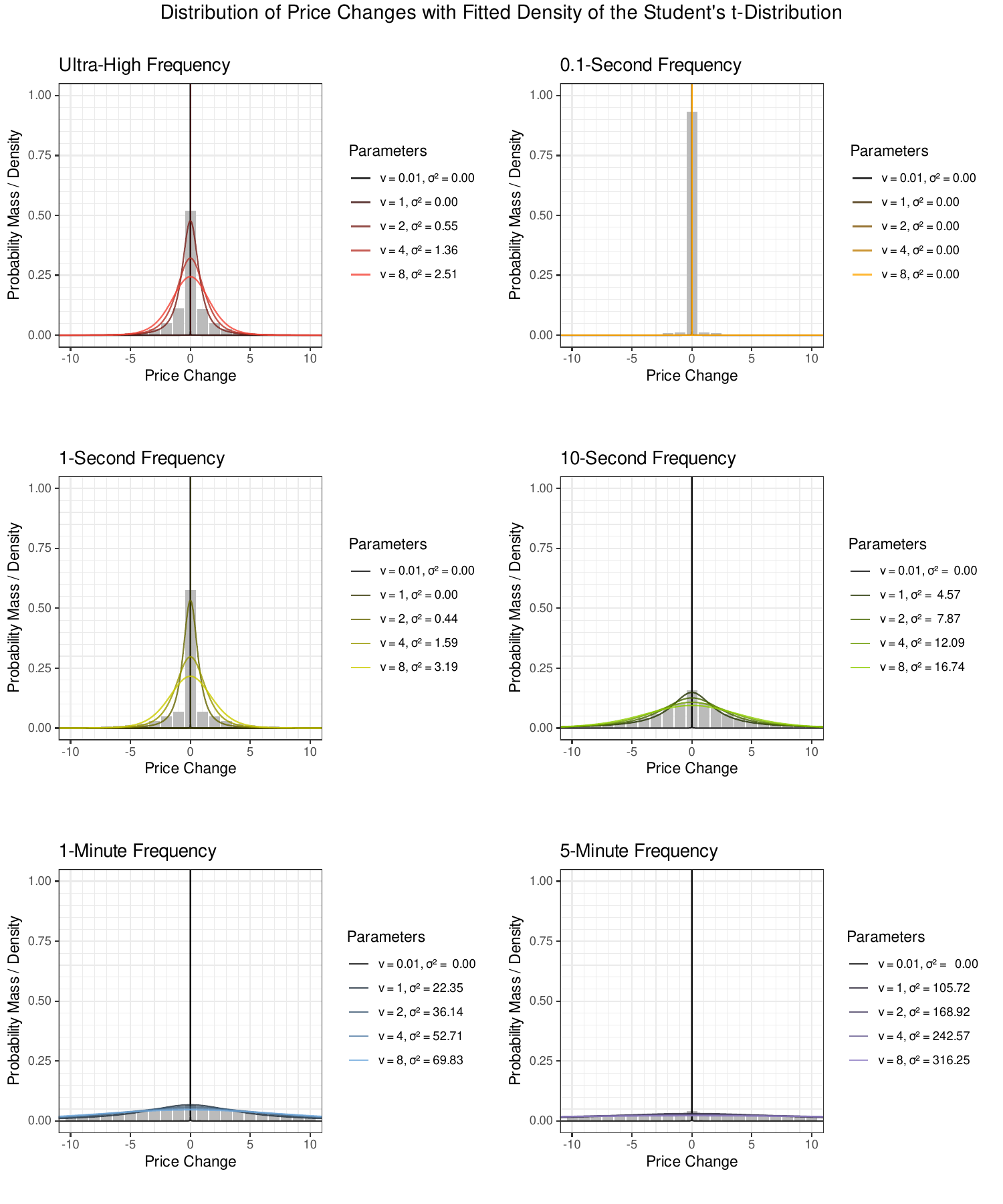}
\caption{The distribution of price changes with the fitted density of the Student's t-distribution, using various fixed degrees of freedom and estimated scale parameters.}
\label{fig:returns}
\end{figure}

Figure \ref{fig:returns} shows the distribution of price changes. We can see that the vast majority of tick-by-tick price changes are quite small--52 percent are zero, while 98 percent lie between -10 and 10. Aggregating the prices to a 1-second frequency produces a similar distribution of price changes as the original ultra-high-frequency data. The associated data are available at \url{github.com/vladimirholy/continuous-high-frequency-garch}.

\section{Unsuitability of Continuous Models}
\label{sec:cont}

We start our analysis by estimating volatility models based on the Student's t-distribution for each trading day. First, we estimate a standard GARCH model of \cite{Engle1982a} and \cite{Bollerslev1986, Bollerslev1987}, given by
\begin{equation}
\label{eq:garch}
y_t = \mu + e_t, \qquad e_t \sim \mathrm{t} \left(0, \sigma_t^2, \nu \right), \qquad \sigma^2_t = \omega + \alpha e_{t-1}^2 + \varphi \sigma^2_{t-1}.
\end{equation}
Second, we estimate a score-driven model of \cite{Creal2013}, also known as the generalized autoregressive score (GAS) model or dynamic conditional score (DCS) model, with a time-varying volatility parameter, given by
\begin{equation}
\label{eq:gas}
y_t \sim \mathrm{t} \left(\mu, \sigma_t^2, \nu \right), \qquad \ln \sigma^2_t = \omega + \alpha \nabla_{\ln \sigma^2} (y_{t-1}; \mu, \sigma^2_{t-1}, \nu) + \varphi \sigma^2_{t-1},
\end{equation}
where $\nabla_{\ln \sigma^2} (y_{t-1}; \mu, \sigma^2_{t-1}, \nu)$ is the score with respect to $\ln \sigma^2$, given by
\begin{equation}
\label{eq:score}
\nabla_{\ln \sigma^2} \left(y; \mu, \sigma^2, \nu \right) = \frac{\partial \ln f \left( y \middle| \mu, \sigma^2, \nu \right)}{\partial \ln \sigma^2}.
\end{equation}
The main difference between the models is that the score-driven model reflects the shape of the Student's t-distribution, particularly its heavy tails, in the dynamics of volatility through the score term. In the relevant literature, score-driven models are found to be superior both theoretically and empirically; see, e.g., \cite{Blasques2015} and \cite{Blazsek2020}. The dynamics in the score-driven model are also specified in terms of the logarithm of the scale parameter, similarly to the EGARCH model of \cite{Nelson1991}.

For the estimation, we utilize four R packages in total. The standard GARCH model is implemented in the \verb"rugarch" package of \cite{Ghalanos2024} and the \verb"fGarch" package of \cite{Wuertz2024}, while the \verb"GAS" package of \cite{Ardia2019} and the \verb"gasmodel" package of \cite{Holy2025b} provide functionality for score-driven models. We leave most arguments at their default values to showcase the basic use of these packages, especially with regard to the numerical optimizers.

The results of the estimation are reported in Table \ref{tab:cont} for 1-second and 1-minute frequencies. While for the 1-minute frequency the results from all packages are quite similar, with a slightly better fit for the score-driven model, the results for the 1-second frequency show striking differences. The main discrepancies with regard to the parameters lie in the degrees of freedom of the Student's t-distribution. Some packages impose a lower bound on $\nu$ to ensure finite moments. The \verb"rugarch" package has a lower bound of 2.1, the \verb"fGarch" package 2, and the \verb"GAS" package 4. In the \verb"gasmodel" package, $\nu$ is unbounded, i.e., it is only required to be positive. While for the first three packages the estimated values of $\nu$ are close to these bounds, the \verb"gasmodel" package estimates $\nu$ with a median value of 0.220. The models with such small values of $\nu$ yield an extremely high log-likelihood--with a median of 72 per observation, in contrast to about -2 for the other three packages. This behavior suggests a significant issue regarding the use of the Student's t-distribution in this context.

\begin{figure}
\centering
\includegraphics[width=\textwidth]{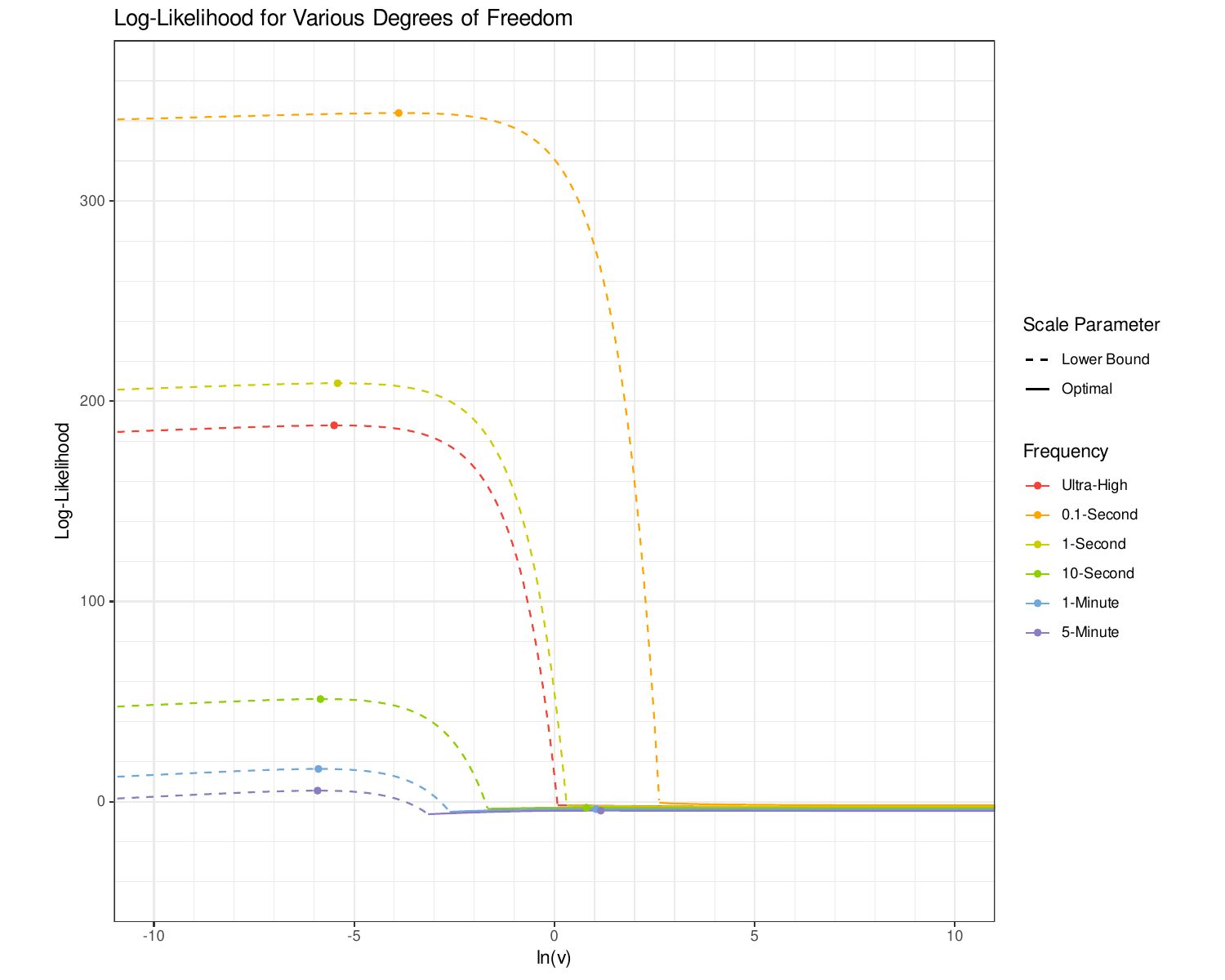}
\caption{The log-likelihood from the fitted Student's t-distribution, using various fixed degrees of freedom and estimated scale parameters. The dashed line represents the scale parameter at the lower bound, $2^{-1074}$, due to numerical precision.}
\label{fig:freedom}
\end{figure}

Next, we focus on a simpler model with the Student's t-distribution, in which $\sigma^2$ is static. Figure \ref{fig:freedom} shows that, for all considered frequencies, the log-likelihood of the static model is maximized when $\nu$ is very close to zero. In these cases, Figure \ref{fig:returns} further shows that the fitted Student's t-distribution degenerates, with $\sigma^2$ approaching zero. The extremely small values of $\sigma^2$ and $\nu$ cause the distribution to be concentrated at 0, with extremely heavy tails. Since the most likely value of price change is zero for all considered frequencies, this concentration of density at 0 leads to an explosion in the log-likelihood. The occurrence of nonzero values is then captured by the heavy tails of the distribution. No moments exist for $\nu \leq 1$. The estimation of the parameters encounters numerical issues. In our case, $\sigma^2$ is estimated as the smallest possible positive double-precision floating-point number, $2^{-1074}$.
Note that, in Table \ref{tab:cont}, when $\sigma^2$ is allowed to be time-varying, the numerical solver in the \verb"gasmodel" package struggles to achieve the optimum, where $\sigma^2$ reaches its lowest possible value and thus remains virtually static. Figures \ref{fig:returns} and \ref{fig:freedom} show that the lower the occurrence of zeros, the lower $\nu$ must be for the log-likelihood and density at 0 to explode.

It follows that the Student's t-distribution is unusable when exact zero values frequently occur. The results from our four packages might appear reasonable at first sight; however, all estimates in Table \ref{tab:cont} are based on suboptimal solutions, either due to artificial bounds on $\nu$ or convergence issues of the numerical solver. Neither the GARCH nor the score-driven models should be relied upon. The optimal solution, where the density takes the shape of $\perp$, is useless for any analysis, as it contains no meaningful information.

\begin{table}
\caption{The median estimated parameters, the ARCH-LM statistic, and the fitted log-likelihood from daily models based on the Student's t-distribution, estimated using various R packages.}
\label{tab:cont}
\centering
\begin{tabular}{lrrrrrrrr}
\toprule
& \multicolumn{4}{c}{1 Second Frequency} & \multicolumn{4}{c}{1 Minute Frequency} \\ \cmidrule(l{3pt}r{3pt}){2-5} \cmidrule(l{3pt}r{3pt}){6-9}
& \verb"rugarch" & \verb"fGarch" & \verb"GAS" & \verb"gasmodel" & \verb"rugarch" & \verb"fGarch" & \verb"GAS" & \verb"gasmodel" \\ 
\midrule
$\mu$ & 0.000 & 0.000 & 0.002 & 0.000 & 0.106 & 0.103 & 0.117 & 0.108 \\ 
$\omega$ & 0.027 & 0.000 & 0.000 & 0.000 & 1.878 & 1.887 & 0.007 & 0.005 \\ 
$\alpha$ & 0.119 & 1.000 & 0.043 & 1.350 & 0.072 & 0.074 & 0.188 & 0.183 \\ 
$\varphi$ & 0.876 & 0.962 & 1.000 & 0.999 & 0.906 & 0.904 & 0.999 & 0.999 \\ \vspace{2mm}
$\nu$ & 2.146 & 2.013 & 4.000 & 0.220 & 6.588 & 6.528 & 6.964 & 6.926 \\
$A$ & 0.001 & 0.002 & 0.022 & x & 0.020 & 0.020 & 0.023 & 0.022 \\ 
$\ell$ & -1.903 & -1.884 & -1.992 & 72.009 & -3.574 & -3.574 & -3.572 & -3.569 \\ 
\bottomrule
\end{tabular}
\begin{flushleft}
\scriptsize
\textit{Notes:}
$\mu$ -- expected price difference; $\omega$ -- intercept in the volatility equation; $\alpha$ -- variance/score coefficient; $\varphi$ -- autoregressive coefficient; $\nu$ -- degrees of freedom; $A$ -- R$^2$ statistic of the ARCH-LM test with lag 10; $\ell$ -- average log-likelihood.
\end{flushleft}
\end{table}

\section{Integer Model and Interval Maximum Likelihood Estimation}
\label{sec:mle}

Fortunately, there is a simple cure for the degenerative behavior of the Student's t-distribution on discrete data with zeros, caused by treating the observations as continuous values and maximizing the log-likelihood in the form of density: The integer observations can be considered as rounded continuous values. For example, instead of the exact value of 2, the observation should be treated as the interval $(-1.5, 2.5]$. In such an approach, the log-likelihood takes the form of the probability of the observed integer $y$ falling into its corresponding interval $(y-0.5, y + 0.5]$. Rounding to the nearest integer is the most natural scheme for our purposes, as it produces intervals of equal length and preserves symmetry. For instance, when 0 is observed, the underlying continuous value is equally likely to have been slightly positive or slightly negative.

Let $F(\cdot| \nu )$ denote the cumulative distribution function of the Student's t-distribution with $\nu$ degrees of freedom. Let $\mu_t$ be the time-varying location parameter and $\sigma^2_t$ the time-varying scale parameter. Let $p$ be the vector of parameters to be estimated, including $\nu$. The interval log-likehood function is then
\begin{equation}
\label{eq:lik}
\ell \left( p \middle| y \right) = \sum_{t=1}^n \ln \left( F \left( \frac{y_t - \mu_t + 0.5}{\sigma_t} \middle | \nu \right) - F \left( \frac{y_t - \mu_t - 0.5}{\sigma_t} \middle | \nu \right) \right).
\end{equation}
Maximizing this function leads to meaningful results, as the discreteness of the data is properly addressed.

We now turn to the specification of our proposed GARCH-like model. We base our model on the score-driven model \eqref{eq:gas}, but modify it in several ways, similarly to \cite{Holy2024d}. The location parameter is made time-varying and follows a first-order moving average process with a long-term value of zero, given by
\begin{equation}
\mu_t = \theta (y_{t-1} - \mu_{t-1}).
\end{equation}
This process captures the market microstructure noise, which occurs at higher frequencies; see, e.g., \cite{Hansen2006} for more details. Along with the empirical evidence for higher frequencies, the intercept is not included.

The scale parameter exhibits diurnal patterns of increased volatility after the market opens and, to a lesser degree, before the market closes. This pattern is estimated using smoothing splines applied to squared price changes and is incorporated into the variable $s_t$, which is then included in the equation for $\sigma^2_t$. The dynamics are captured in a separate component $e_t$ to ensure that the autoregressive part omits the lagged value $s_{t-1}$. Separating $\omega$ and $\varphi$ also accelerates the convergence of the numerical optimization. The scale parameter then follows
\begin{equation}
\ln \sigma^2_t = \omega + \ln \hat{s}_t + e_t, \qquad e_t = \alpha \nabla_{\ln \sigma^2} \left(y_{t-1}; \mu_{t-1}, \sigma^2_{t-1}, \nu \right) + \varphi e_{t-1}.
\end{equation}
The score accounts for the interval structure, just as the interval log-likelihood \eqref{eq:lik}, and takes the form
\begin{equation}
\nabla_{\ln \sigma^2} \left(y; \mu, \sigma^2, \nu \right) = \frac{\left( y - \mu - 0.5 \right) f \left( \frac{y - \mu - 0.5}{\sigma} \middle | \nu \right) - \left( y - \mu + 0.5 \right) f \left( \frac{y - \mu + 0.5}{\sigma} \middle | \nu \right)}{2 \sigma F \left( \frac{y - \mu + 0.5}{\sigma} \middle | \nu \right) - 2 \sigma F \left( \frac{y - \mu - 0.5}{\sigma} \middle | \nu \right)},
\end{equation}
where $f(\cdot| \nu )$ denotes the density function of the Student's t-distribution.

\section{Performance of Integer Models}
\label{sec:integer}

We estimate the proposed model using 1-second and 1-minute frequency data. We compare it with three models that have equivalent dynamics but different distributions: the normal distribution with our interval-based estimation approach, the discrete Skellam distribution, and its zero-inflated version with an additional zero-inflation parameter $\pi$. For the Skellam distributions, we model the time-varying overdispersion parameter as proposed by \cite{Holy2024d}.

\begin{figure}
\centering
\includegraphics[width=\textwidth]{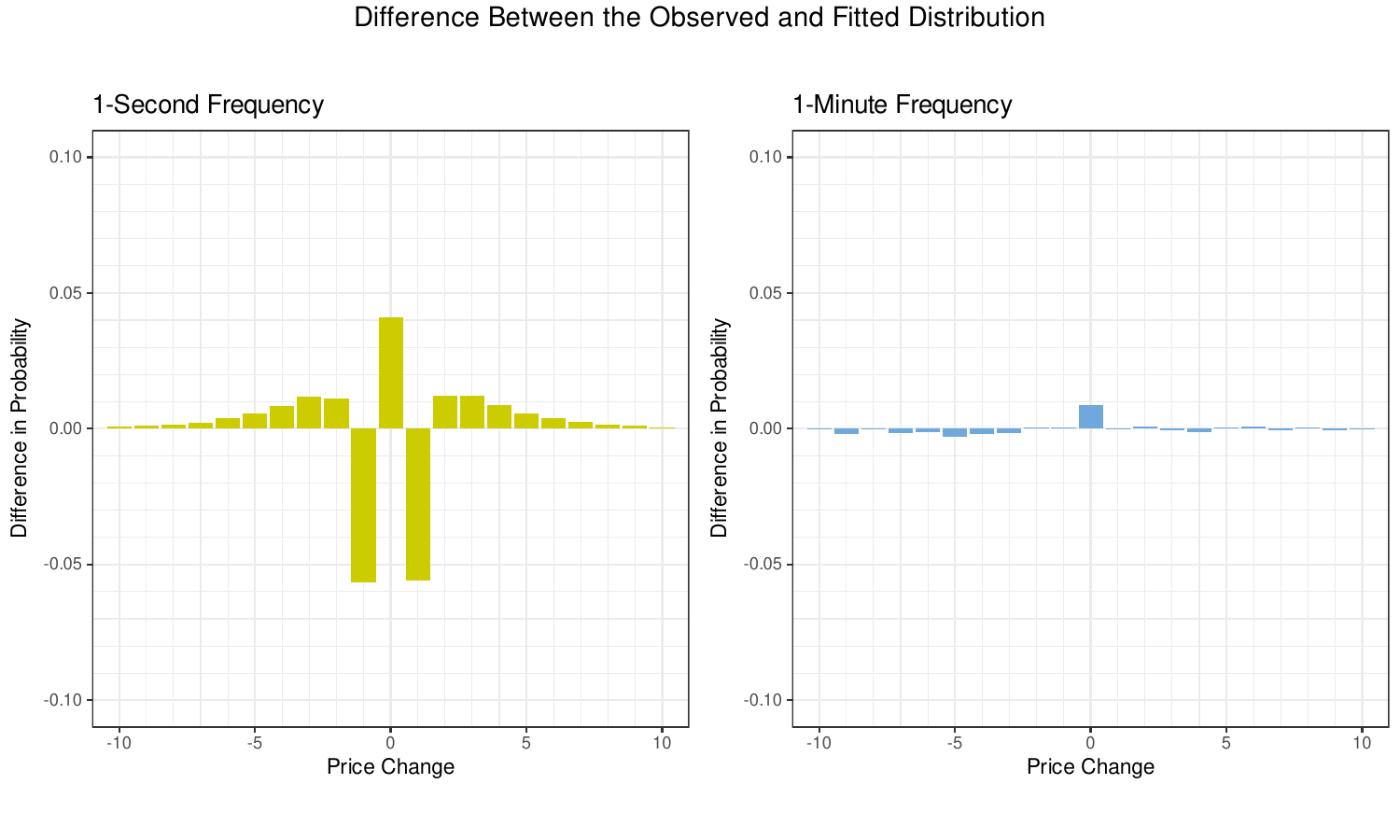}
\caption{The average difference between the observed distribution of price changes and the fitted probabilities from the daily integer models based on the Student's t-distribution.}
\label{fig:fit}
\end{figure}

The results are reported in Table \ref{tab:integer}. Note that the reported log-likelihood $\ell$ is based on probability mass functions, in contrast to the density functions in Table \ref{tab:cont}. We observe that for the 1-second frequency, the zero-inflated Skellam model performs best, while for the 1-minute frequency, the integer Student's t-model provides a better fit. Figure \ref{fig:fit} shows that the Student's t-distribution is not flexible enough to capture price changes in 1-second frequency data, with the probabilities of -1 and 1 being overestimated and the remaining probabilities being underestimated. For 1-minute frequency, there does not appear to be a systematic bias, although the probability of 0 is underestimated.

To detect the presence of ARCH effects in the residuals, Table \ref{tab:integer} also reports the R$^2$ statistic from the ARCH-LM test of \cite{Engle1982a}, with the maximum lag set to 10. For all models, autocorrelation in squared residuals is quite low, indicating that they capture time-varying volatility well, except for the model based on the Student's t-distribution with 1 second frequency, for which standardized residuals cannot be obtained because the first and second moments do not exist for most days.

\begin{table}
\caption{The median estimated parameters, the ARCH-LM statistic, and the fitted log-likelihood from daily models based on various integer distributions.}
\label{tab:integer}
\centering
\begin{tabular}{lrrrrrrrr}
\toprule
& \multicolumn{4}{c}{1 Second Frequency} & \multicolumn{4}{c}{1 Minute Frequency} \\ \cmidrule(l{3pt}r{3pt}){2-5} \cmidrule(l{3pt}r{3pt}){6-9}
& Normal & t & Skellam & Z-I Sk.& Normal & t & Skellam & Z-I Sk. \\ 
\midrule
$\theta$ & -0.291 & -0.028 & -0.328 & -0.612 & -0.055 & -0.059 & -0.059 & -0.056 \\ 
$\omega$ & 2.559 & -1.378 & 1.721 & 2.432 & 4.585 & 4.333 & 4.351 & 4.345 \\ 
$\alpha$ & 0.040 & 0.084 & 0.045 & 0.074 & 0.083 & 0.113 & 0.043 & 0.042 \\ 
$\varphi$ & 0.977 & 0.998 & 0.995 & 0.994 & 0.892 & 0.915 & 0.990 & 0.991 \\ 
$\nu$ &  & 0.908 &  &   &  & 9.268 &  & \\ \vspace{2mm}
$\pi$ &  &  &  & 0.492 &  &  &  & 0.011 \\ 
$A$ & 0.004 & x & 0.001 & 0.001 & 0.018 & 0.019 & 0.021 & 0.022 \\ \vspace{2mm}
$\ell$ & -2.228 & -1.841 & -2.068 & -1.700 & -3.565 & -3.550 & -3.566 & -3.557 \\ 
$A_\text{F}$ & x & x & 0.002 & 0.001 & 0.024 & 0.026 & 0.025 & 0.026 \\ 
$\ell_\text{F}$ & x & -1.844 & -2.144 & -1.787 & -3.644 & -3.629 & -3.779 & -3.903 \\ 
\bottomrule
\end{tabular}
\begin{flushleft}
\scriptsize
\textit{Notes:}
$\theta$ -- moving-average coefficient; $\omega$ -- intercept in the volatility equation; $\alpha$ -- score coefficient; $\varphi$ -- autoregressive coefficient; $\nu$ -- degrees of freedom; $\pi$ -- zero inflation; $A$, $A_\text{F}$ -- R$^2$ statistic of the ARCH-LM test with lag 10; $\ell$, $\ell_\text{F}$ -- average log-likelihood; $A_\text{F}$, $\ell_\text{F}$ are evaluated on the data from the next day.
\end{flushleft}
\end{table}

Finally, we assess out-of-sample performance. For this purpose, we evaluate the daily models on the day following their estimation. For the Student's t, Skellam, and zero-inflated Skellam distributions, Table \ref{tab:integer} shows that the out-of-sample log-likelihoods are only slightly worse than their in-sample counterparts, suggesting stable behavior on a day-to-day basis. The normal distribution at 1-second frequency, however, fails to capture out-of-sample outliers on 56 percent of days, resulting in numerically zero likelihood, and is therefore unsuitable for forecasting studies. Overall, the out-of-sample results confirm the zero-inflated Skellam model as the most suitable for the 1-second frequency, and the Student's t-distribution for the 1-minute frequency.

\section{Conclusion}
\label{sec:con}

The Student's t-distribution is unsuitable for modeling high-frequency integer price changes, as its density function degenerates into a $\perp$-shaped form concentrated at 0 with extremely heavy tails, providing no useful information about the data. However, we show that maximum likelihood estimation can be modified to treat these integer observations as being produced by rounding and replace them with corresponding intervals. The score-driven model with dynamic volatility based on the Student's t-distribution, estimated using this interval approach, produces meaningful results comparable to dynamic models based on the Skellam distribution, which have been proposed in the recent literature. Our empirical analysis shows that the Student's t-distribution is suitable for relatively lower frequencies, such as 1-minute data, but is insufficient to capture the distribution of price changes at relatively higher frequencies. However, our interval approach is in no way restricted to the Student's t-distribution and can be applied to any continuous distribution. This opens up possibilities for future research to consider not only discrete distributions but also continuous distributions for modeling high-frequency prices.

\section*{Funding}
\label{sec:fund}

The work on this paper was supported by the Czech Science Foundation under project 23-06139S and the personal and professional development support program of the Faculty of Informatics and Statistics, Prague University of Economics and Business.


\appendix

\section{Evidence from Further Stocks}
\label{app:further}

\begin{figure}
\centering
\includegraphics[width=\textwidth]{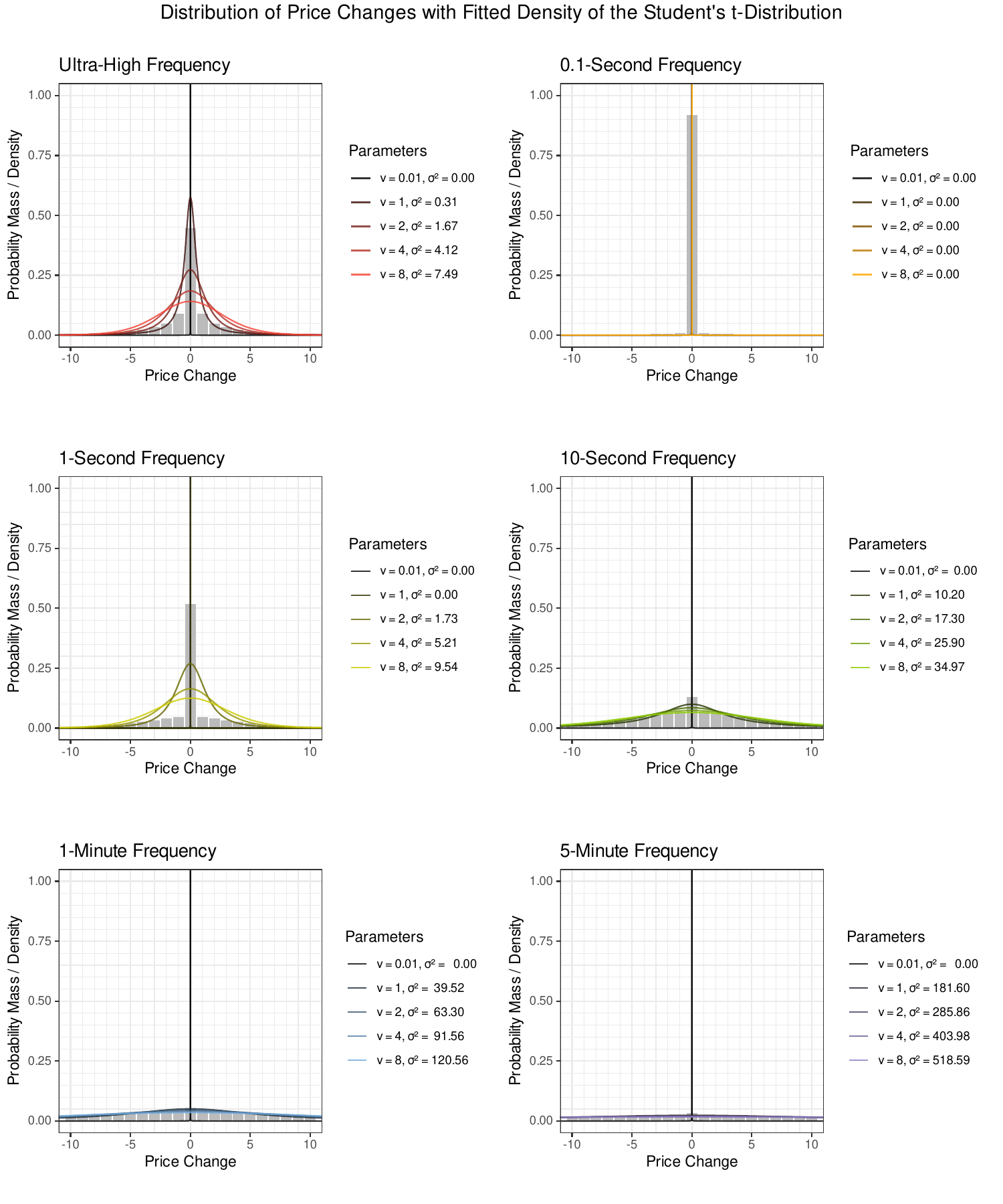}
\caption{\textbf{The MCD stock} -- The distribution of price changes with the fitted density of the Student's t-distribution, using various fixed degrees of freedom and estimated scale parameters.}
\label{fig:returnsMCD}
\end{figure}

\begin{figure}
\centering
\includegraphics[width=\textwidth]{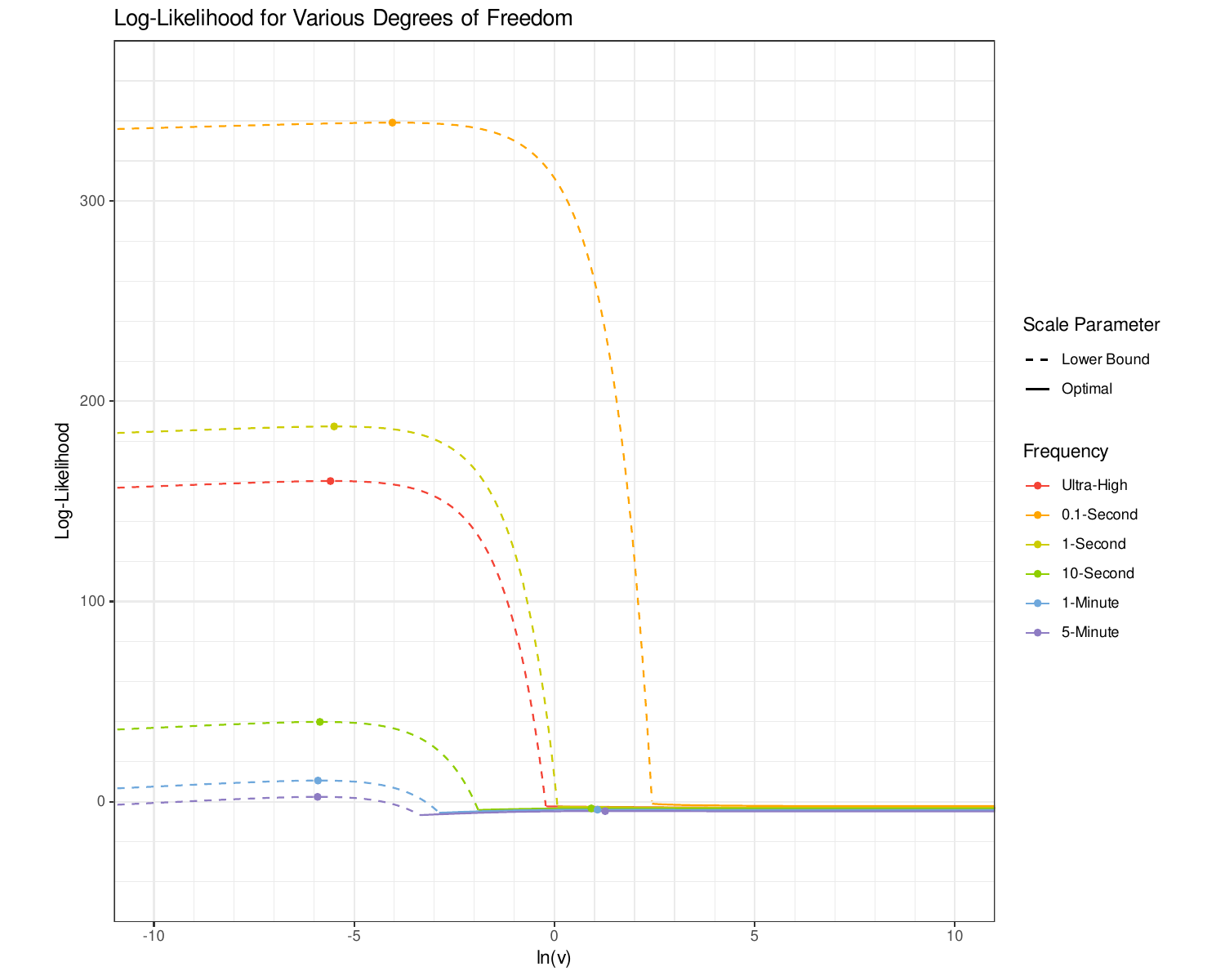}
\caption{\textbf{The MCD stock} -- The log-likelihood from the fitted Student's t-distribution, using various fixed degrees of freedom and estimated scale parameters. The dashed line represents the scale parameter at the lower bound, $2^{-1074}$, due to numerical precision.}
\label{fig:freedomMCD}
\end{figure}

\begin{figure}
\centering
\includegraphics[width=\textwidth]{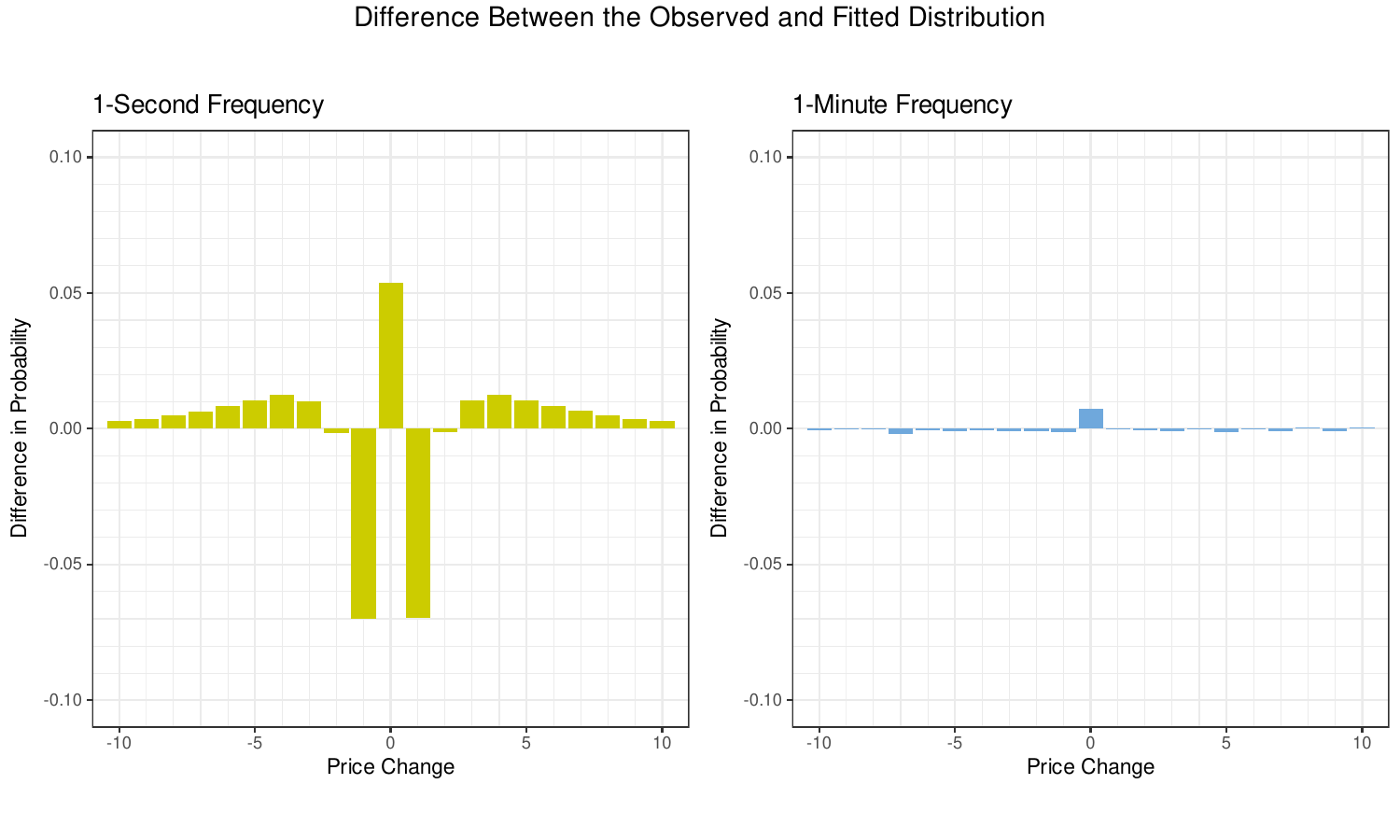}
\caption{\textbf{The MCD stock} -- The average difference between the observed distribution of price changes and the fitted probabilities from the daily integer models based on the Student's t-distribution.}
\label{fig:fitMCD}
\end{figure}

\begin{figure}
\centering
\includegraphics[width=\textwidth]{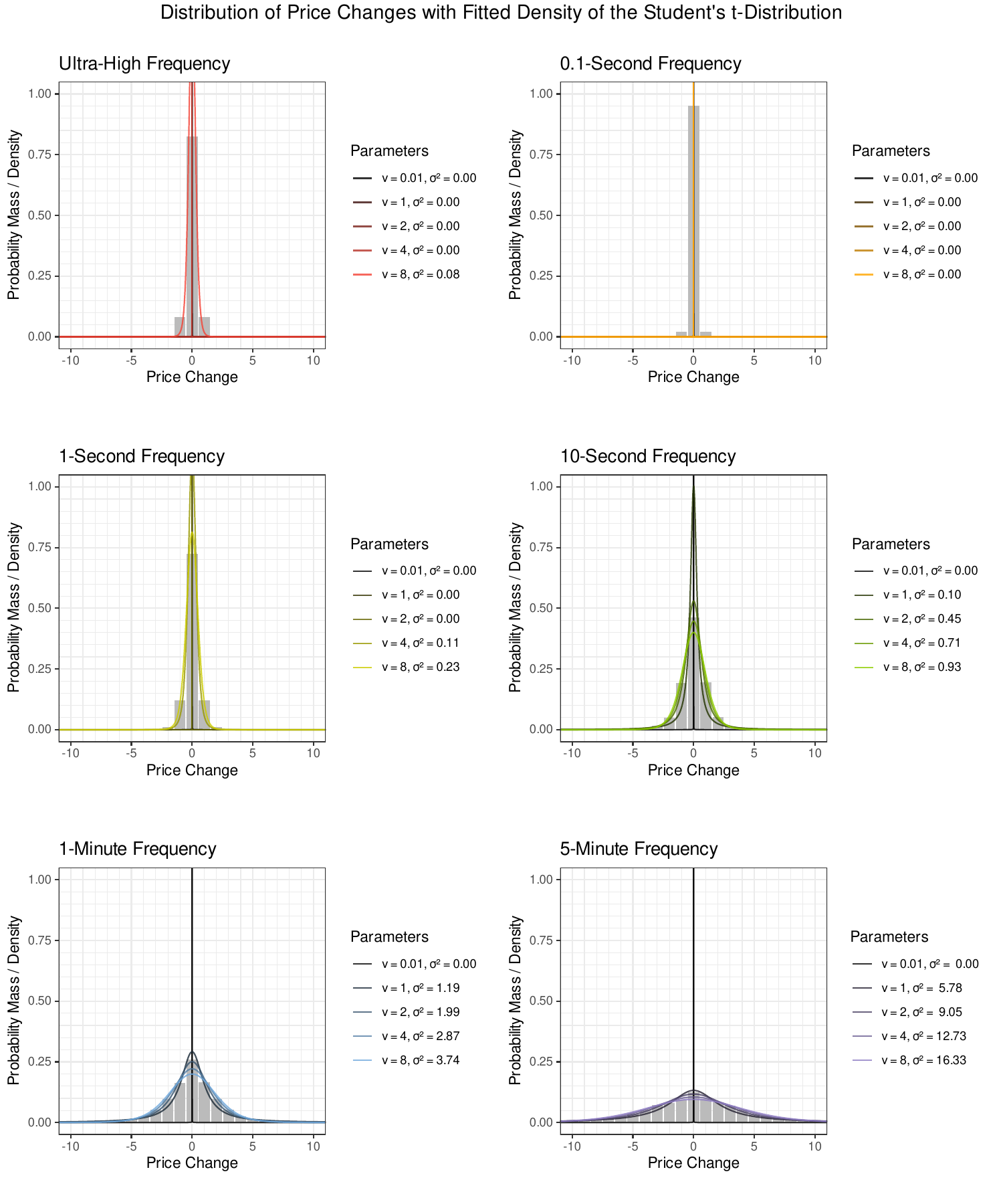}
\caption{\textbf{The CSCO stock} -- The distribution of price changes with the fitted density of the Student's t-distribution, using various fixed degrees of freedom and estimated scale parameters.}
\label{fig:returnsCSCO}
\end{figure}

\begin{figure}
\centering
\includegraphics[width=\textwidth]{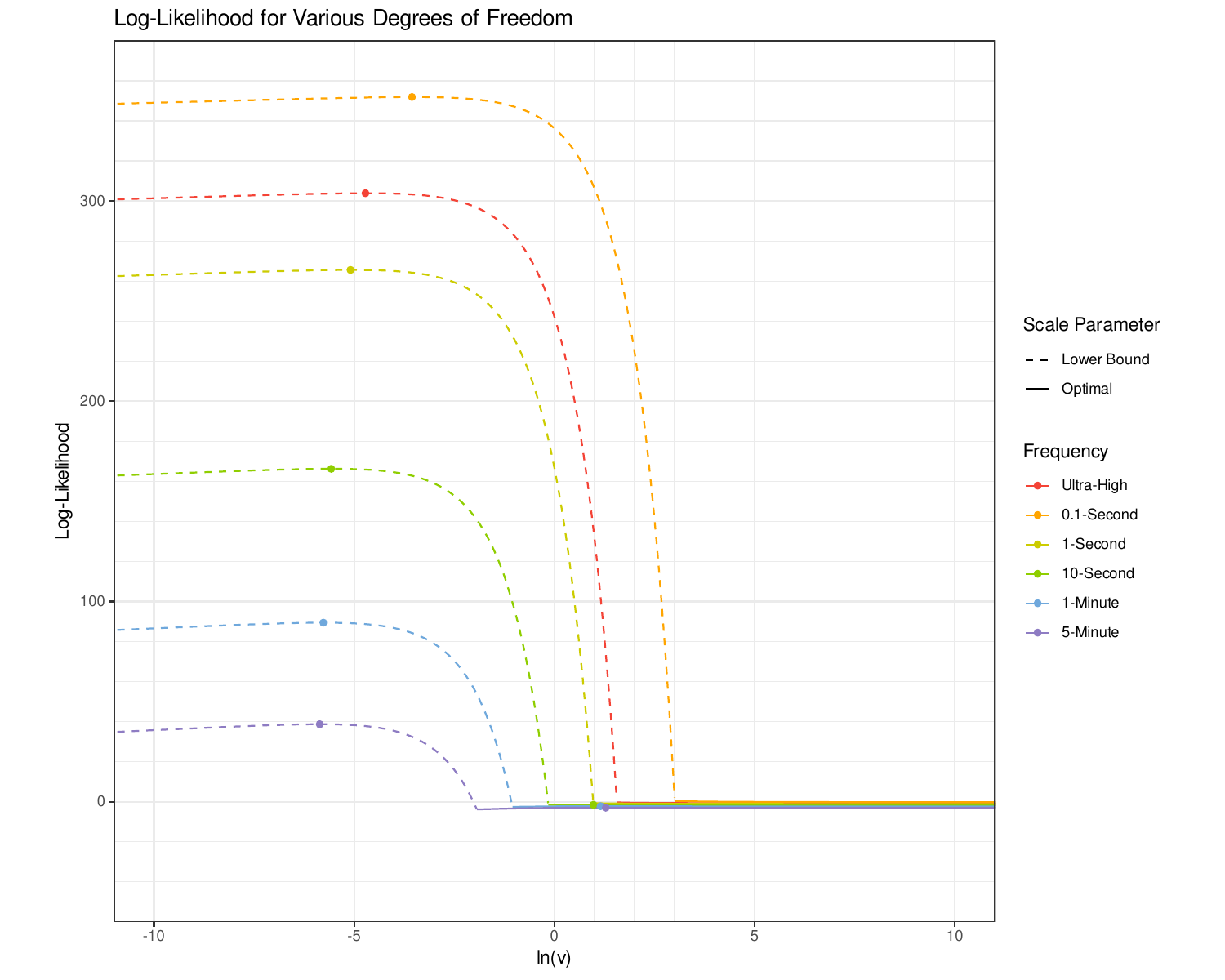}
\caption{\textbf{The CSCO stock} -- The log-likelihood from the fitted Student's t-distribution, using various fixed degrees of freedom and estimated scale parameters. The dashed line represents the scale parameter at the lower bound, $2^{-1074}$, due to numerical precision.}
\label{fig:freedomCSCO}
\end{figure}

\begin{figure}
\centering
\includegraphics[width=\textwidth]{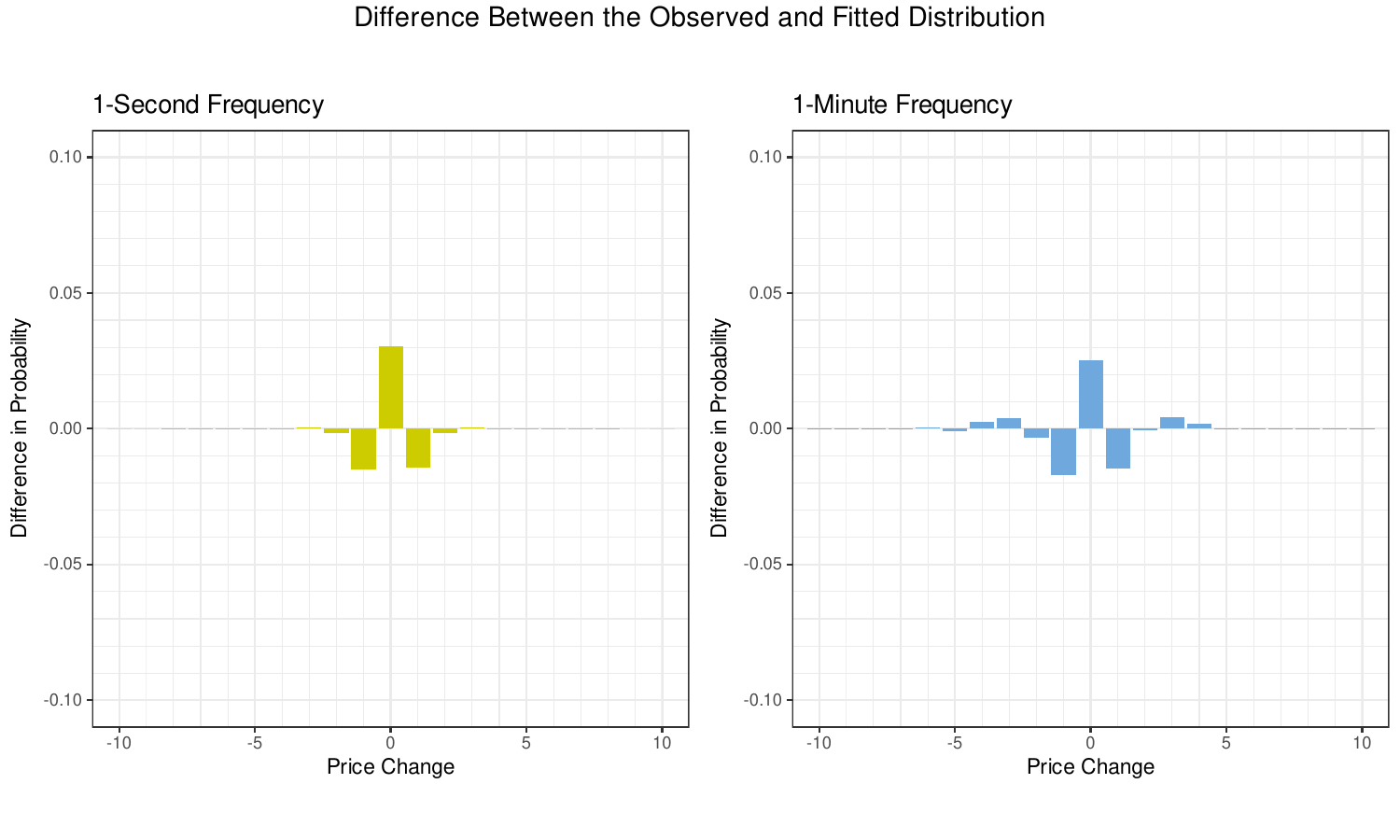}
\caption{\textbf{The CSCO stock} -- The average difference between the observed distribution of price changes and the fitted probabilities from the daily integer models based on the Student's t-distribution.}
\label{fig:fitCSCO}
\end{figure}

\begin{figure}
\centering
\includegraphics[width=\textwidth]{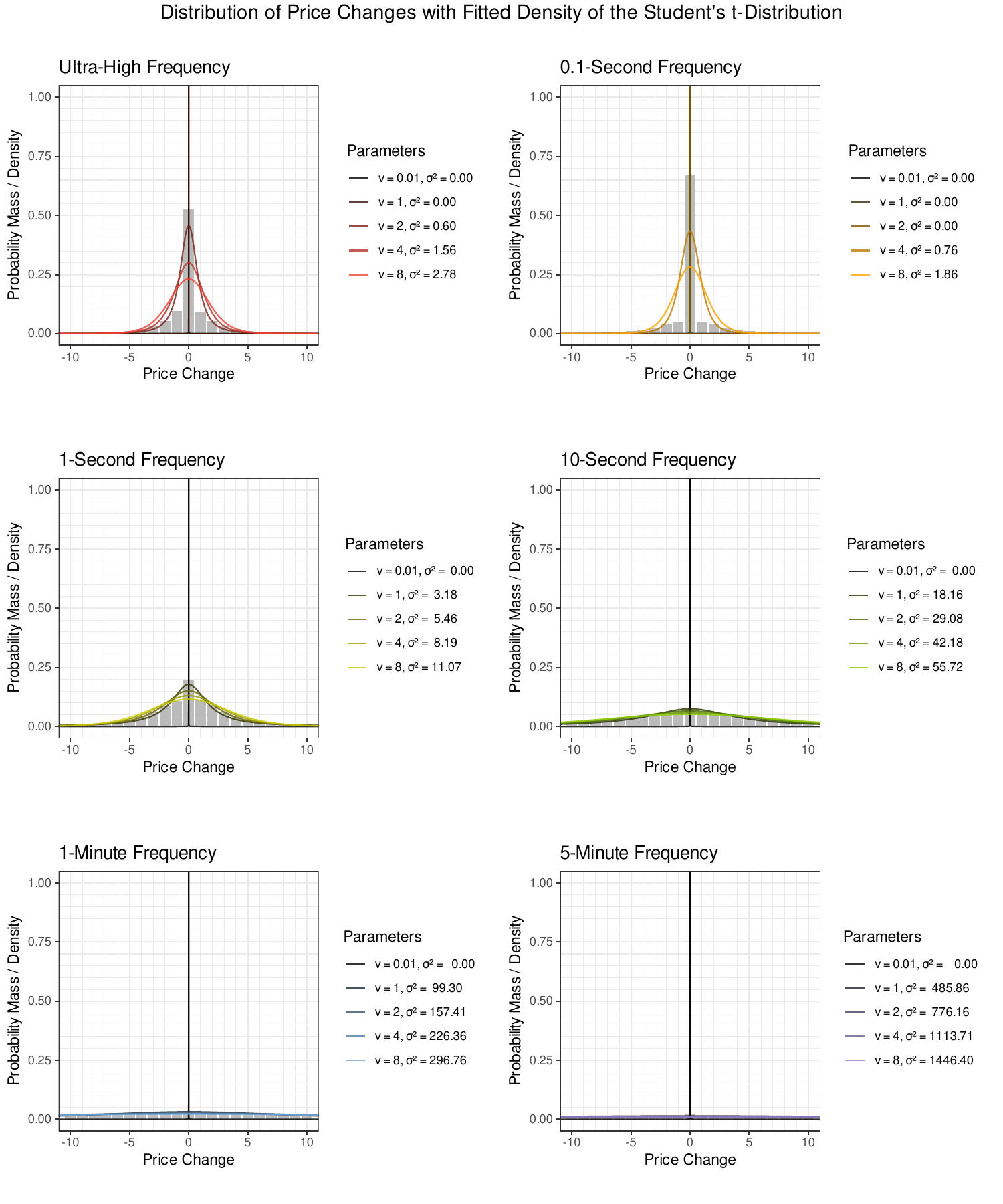}
\caption{\textbf{The MSFT stock} -- The distribution of price changes with the fitted density of the Student's t-distribution, using various fixed degrees of freedom and estimated scale parameters.}
\label{fig:returnsMSFT}
\end{figure}

\begin{figure}
\centering
\includegraphics[width=\textwidth]{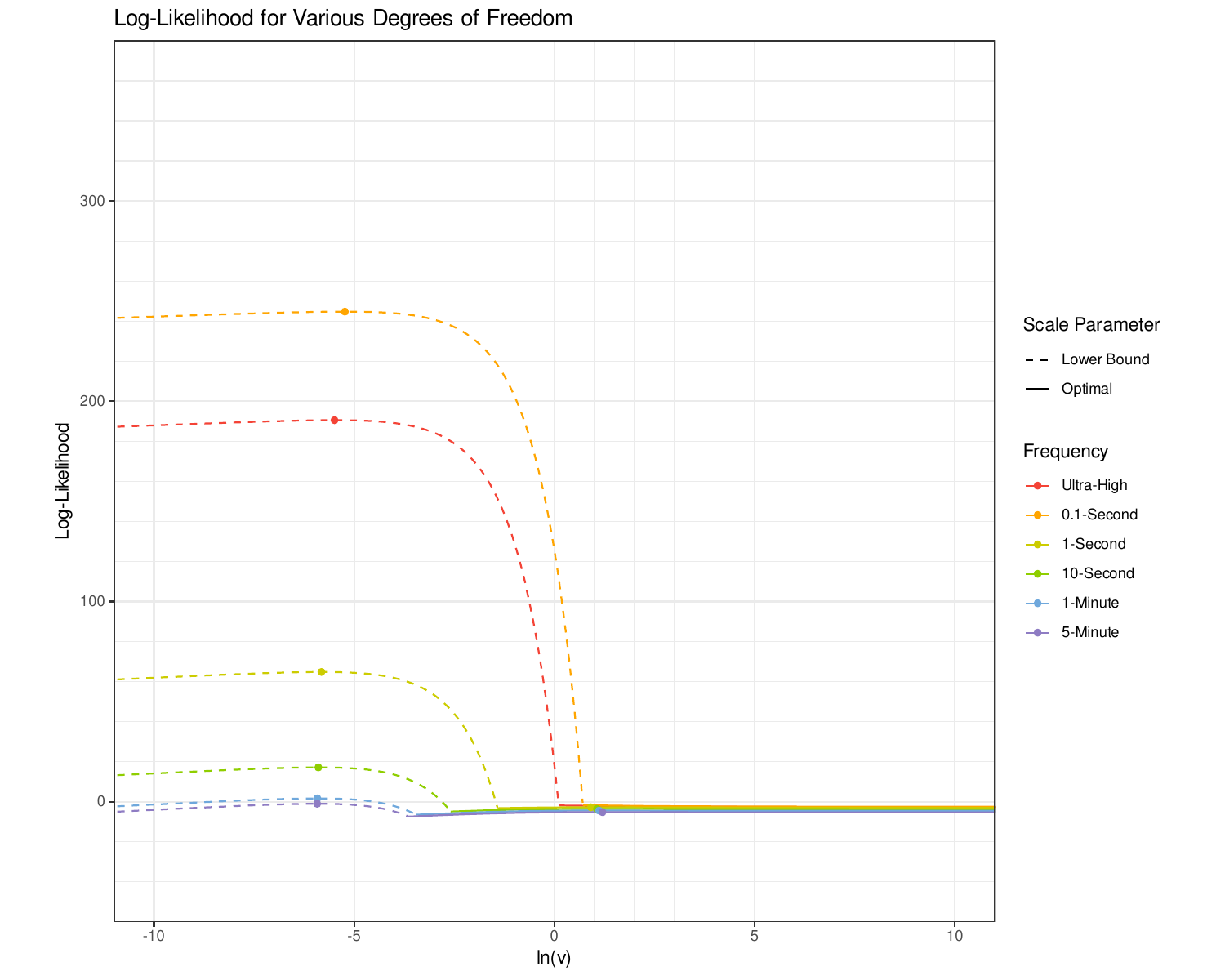}
\caption{\textbf{The MSFT stock} -- The log-likelihood from the fitted Student's t-distribution, using various fixed degrees of freedom and estimated scale parameters. The dashed line represents the scale parameter at the lower bound, $2^{-1074}$, due to numerical precision.}
\label{fig:freedomMSFT}
\end{figure}

\begin{figure}
\centering
\includegraphics[width=\textwidth]{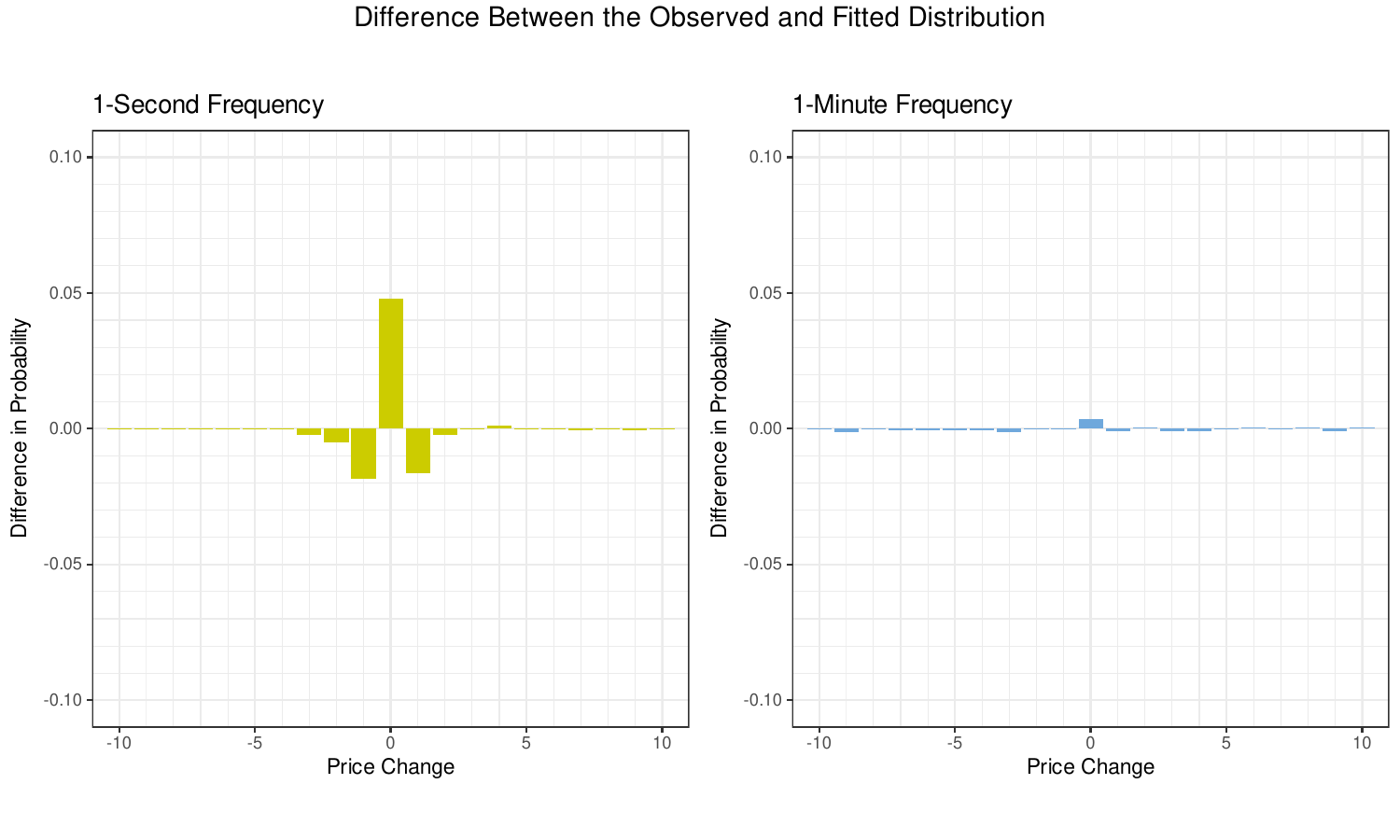}
\caption{\textbf{The MSFT stock} -- The average difference between the observed distribution of price changes and the fitted probabilities from the daily integer models based on the Student's t-distribution.}
\label{fig:fitMSFT}
\end{figure}

To provide more general insights, we further analyze the MCD, CSCO, and MSFT stocks in addition to the IBM stock. This set of four stocks has also been analyzed, for example, by \cite{Blasques2024a}. All four stocks are components of the Dow Jones Industrial Average (DJIA). The main results regarding the degenerative behavior of the Student's t-distribution observed for the IBM stock are also found for these three stocks. Similarly, we also observe limitations of some models in forecasting. We present the results in Figures \ref{fig:returnsMCD}--\ref{fig:fitMSFT} and Tables \ref{tab:contMCD}--\ref{tab:integerMSFT}, and discuss differences in behavior from the IBM stock below.

First, we focus on the MCD stock of McDonald’s Corporation traded on the NYSE. The number of tick observations after data cleaning is quite similar to that of IBM, with over 15 million observations. Figures \ref{fig:returnsMCD}--\ref{fig:fitMCD} and Table \ref{tab:contMCD} correspond to Figures \ref{fig:returns}--\ref{fig:fit} and Table \ref{tab:cont}. In Table \ref{tab:integerMCD}, there is one notable difference: for the 1-minute frequency, the score coefficient is estimated as negative on most days for the normal distribution. These negative values lack a clear interpretation and are a product of overfitting, as evidenced by the out-of-sample analysis, where the model performs quite poorly.

Second, we analyze the CSCO stock of Cisco Systems, Inc., traded on the NASDAQ. The number of tick observations is almost 27 million. There is a much larger occurrence of zero values than for the IBM stock, as illustrated in Figure \ref{fig:returnsCSCO}. Figure \ref{fig:freedomCSCO} and Table \ref{tab:contCSCO} correspond to Figure \ref{fig:freedom} and Table \ref{tab:cont}. Table \ref{tab:integerCSCO} shows that the estimated degrees of freedom of the Student's t-distribution at the 1-second frequency are larger than for the IBM stock. Similar to the MCD stock, negative values of the score coefficient occur at the 1-minute frequency for the normal, Skellam, and zero-inflated Skellam distributions. The Student’s t-distribution maintains positive values but exhibits a slightly worse in-sample fit and a dramatically better out-of-sample fit. This suggests that it could be practical to impose a lower bound of zero on the score coefficient. Figure \ref{fig:fitCSCO} shows systematic biases at both the 1-second and 1-minute frequencies.

Third, we analyze the MSFT stock of Microsoft Corporation traded on the NASDAQ. The number of tick observations is the highest among the studied stocks, with about 90 million observations. The distributions in Figure \ref{fig:returnsMSFT} are more spread out than those for the IBM stock. Consequently, Table \ref{tab:contMSFT} reports more stable results across the four R packages, with degrees of freedom estimated above 4 by all packages for the 1-second frequency. Nevertheless, these are all suboptimal solutions, as evident from Figure \ref{fig:freedomMSFT}. Table \ref{tab:integerMSFT} shows that at the 1-second frequency, the Student's t-distribution performs best. At the 1-minute frequency, it is surpassed by the normal distribution with a negative score coefficient in-sample, but the Student's t-distribution outperforms out-of-sample. The Skellam distribution, along with its zero-inflated version, is unable to capture out-of-sample values that can be quite spread out on some days, resulting in zero likelihood. Figure \ref{fig:fitMSFT} is consistent with Figure \ref{fig:fit}.

\begin{table}
\caption{\textbf{The MCD stock} -- The median estimated parameters, the ARCH-LM statistic, and the fitted log-likelihood from daily models based on the Student's t-distribution, estimated using various R packages.}
\label{tab:contMCD}
\centering
\begin{tabular}{lrrrrrrrr}
\toprule
& \multicolumn{4}{c}{1 Second Frequency} & \multicolumn{4}{c}{1 Minute Frequency} \\ \cmidrule(l{3pt}r{3pt}){2-5} \cmidrule(l{3pt}r{3pt}){6-9}
& \verb"rugarch" & \verb"fGarch" & \verb"GAS" & \verb"gasmodel" & \verb"rugarch" & \verb"fGarch" & \verb"GAS" & \verb"gasmodel" \\ 
\midrule
$\mu$ & 0.002 & 0.002 & 0.003 & -0.000 & 0.097 & 0.085 & 0.089 & 0.088 \\ 
$\omega$ & 0.556 & 0.000 & 0.001 & -0.001 & 3.362 & 3.369 & 0.007 & 0.000 \\ 
$\alpha$ & 0.123 & 1.000 & 0.062 & 2.579 & 0.063 & 0.065 & 0.179 & 0.176 \\ 
$\varphi$ & 0.876 & 0.963 & 1.000 & 0.999 & 0.907 & 0.906 & 0.999 & 1.000 \\ \vspace{2mm}
$\nu$ & 2.213 & 2.015 & 4.000 & 0.201 & 6.150 & 6.103 & 6.981 & 7.007 \\ 
$A$ & 0.003 & 0.004 & 0.022 & x & 0.020 & 0.020 & 0.022 & 0.022 \\ 
$\ell$ & -2.514 & -2.496 & -2.548 & 41.470 & -3.862 & -3.862 & -3.861 & -3.857 \\ 
\bottomrule
\end{tabular}
\begin{flushleft}
\scriptsize
\textit{Notes:}
$\mu$ -- expected price difference; $\omega$ -- intercept in the volatility equation; $\alpha$ -- variance/score coefficient; $\varphi$ -- autoregressive coefficient; $\nu$ -- degrees of freedom; $A$ -- R$^2$ statistic of the ARCH-LM test with lag 10; $\ell$ -- average log-likelihood.
\end{flushleft}
\end{table}

\begin{table}
\caption{\textbf{The MCD stock} -- The median estimated parameters, the ARCH-LM statistic, and the fitted log-likelihood from daily models based on various integer distributions.}
\label{tab:integerMCD}
\centering
\begin{tabular}{lrrrrrrrr}
\toprule
& \multicolumn{4}{c}{1 Second Frequency} & \multicolumn{4}{c}{1 Minute Frequency} \\ \cmidrule(l{3pt}r{3pt}){2-5} \cmidrule(l{3pt}r{3pt}){6-9}
& Normal & t & Skellam & Z-I Sk.& Normal & t & Skellam & Z-I Sk. \\ 
\midrule
$\theta$ & -0.467 & -0.014 & -0.481 & -0.721 & -0.073 & -0.067 & -0.075 & -0.073 \\
$\omega$ & 3.059 & -1.453 & 2.596 & 2.915 & 5.147 & 4.898 & 4.376 & 4.357 \\
$\alpha$ & 0.073 & 0.094 & 0.053 & 0.057 & -0.074 & 0.067 & 0.035 & 0.035 \\
$\varphi$ & 0.980 & 0.999 & 0.991 & 0.995 & 0.972 & 0.946 & 1.000 & 1.000 \\
$\nu$ &  & 0.674 &  &  &  & 9.161 &  &  \\ \vspace{2mm}
$\pi$ &  &  &  & 0.474 &  &  &  & 0.012 \\
$A$ & 0.003 & x & 0.002 & 0.001 & 0.023 & 0.023 & 0.024 & 0.024 \\ \vspace{2mm}
$\ell$ & -2.668 & -2.344 & -2.570 & -2.044 & -3.821 & -3.821 & -3.870 & -3.860 \\
$A_\text{F}$ & x & x & 0.003 & 0.002 & 0.034 & 0.033 & 0.030 & 0.029 \\
$\ell_\text{F}$ & x & -2.365 & -2.955 & -2.373 & -5.228 & -4.101 & -4.297 & -4.441 \\
\bottomrule
\end{tabular}
\begin{flushleft}
\scriptsize
\textit{Notes:}
$\theta$ -- moving-average coefficient; $\omega$ -- intercept in the volatility equation; $\alpha$ -- score coefficient; $\varphi$ -- autoregressive coefficient; $\nu$ -- degrees of freedom; $\pi$ -- zero inflation; $A$, $A_\text{F}$ -- R$^2$ statistic of the ARCH-LM test with lag 10; $\ell$, $\ell_\text{F}$ -- average log-likelihood; $A_\text{F}$, $\ell_\text{F}$ are evaluated on the data from the next day.
\end{flushleft}
\end{table}

\begin{table}
\caption{\textbf{The CSCO stock} -- The median estimated parameters, the ARCH-LM statistic, and the fitted log-likelihood from daily models based on the Student's t-distribution, estimated using various R packages.}
\label{tab:contCSCO}
\centering
\begin{tabular}{lrrrrrrrr}
\toprule
& \multicolumn{4}{c}{1 Second Frequency} & \multicolumn{4}{c}{1 Minute Frequency} \\ \cmidrule(l{3pt}r{3pt}){2-5} \cmidrule(l{3pt}r{3pt}){6-9}
& \verb"rugarch" & \verb"fGarch" & \verb"GAS" & \verb"gasmodel" & \verb"rugarch" & \verb"fGarch" & \verb"GAS" & \verb"gasmodel" \\ 
\midrule
$\mu$ & -0.000 & 0.000 & 0.000 & -0.000 & 0.002 & 0.000 & 0.005 & 0.000 \\ 
$\omega$ & 0.000 & 0.000 & -0.498 & -0.004 & 0.113 & 0.113 & 0.003 & 0.000 \\ 
$\alpha$ & 0.013 & 0.516 & 1.115 & 2.314 & 0.059 & 0.060 & 0.177 & 0.173 \\ 
$\varphi$ & 0.013 & 0.004 & 0.833 & 0.991 & 0.914 & 0.912 & 0.999 & 0.999 \\ \vspace{2mm}
$\nu$ & 2.103 & 2.000 & 4.000 & 0.296 & 6.122 & 6.065 & 6.393 & 6.373 \\ 
$A$ & 0.001 & 0.004 & 0.004 & x & 0.021 & 0.021 & 0.024 & 0.022 \\ 
$\ell$ & 1.887 & 1.876 & -0.718 & 45.858 & -2.130 & -2.130 & -2.132 & -2.126 \\ 
\bottomrule
\end{tabular}
\begin{flushleft}
\scriptsize
\textit{Notes:}
$\mu$ -- expected price difference; $\omega$ -- intercept in the volatility equation; $\alpha$ -- variance/score coefficient; $\varphi$ -- autoregressive coefficient; $\nu$ -- degrees of freedom; $A$ -- R$^2$ statistic of the ARCH-LM test with lag 10; $\ell$ -- average log-likelihood.
\end{flushleft}
\end{table}

\begin{table}
\caption{\textbf{The CSCO stock} -- The median estimated parameters, the ARCH-LM statistic, and the fitted log-likelihood from daily models based on various integer distributions.}
\label{tab:integerCSCO}
\centering
\begin{tabular}{lrrrrrrrr}
\toprule
& \multicolumn{4}{c}{1 Second Frequency} & \multicolumn{4}{c}{1 Minute Frequency} \\ \cmidrule(l{3pt}r{3pt}){2-5} \cmidrule(l{3pt}r{3pt}){6-9}
& Normal & t & Skellam & Z-I Sk.& Normal & t & Skellam & Z-I Sk. \\ 
\midrule
$\theta$ & -0.491 & -0.404 & -0.490 & -0.660 & -0.092 & -0.093 & -0.078 & -0.087 \\
$\omega$ & -0.634 & -1.897 & -1.557 & -1.217 & 1.586 & 1.337 & 1.573 & 1.621 \\
$\alpha$ & 0.127 & 0.494 & 0.038 & 0.073 & -0.034 & 0.034 & -0.120 & -0.098 \\
$\varphi$ & 0.953 & 0.736 & 0.994 & 0.992 & 0.981 & 0.963 & 0.982 & 0.988 \\
$\nu$ &  & 3.717 &  &  &  & 8.505 &  &  \\ \vspace{2mm}
$\pi$ &  &  &  & 0.290 &  &  &  & 0.032 \\
$A$ & 0.004 & 0.005 & 0.002 & 0.001 & 0.024 & 0.024 & 0.024 & 0.024 \\ \vspace{2mm}
$\ell$ & -0.962 & -0.840 & -0.841 & -0.830 & -2.093 & -2.093 & -2.086 & -2.082 \\
$A_\text{F}$ & x & 0.006 & 0.002 & 0.001 & 0.037 & 0.035 & 0.035 & 0.032 \\
$\ell_\text{F}$ & x & -0.843 & -0.843 & -0.831 & -3.323 & -2.360 & -4.956 & -5.506 \\
\bottomrule
\end{tabular}
\begin{flushleft}
\scriptsize
\textit{Notes:}
$\theta$ -- moving-average coefficient; $\omega$ -- intercept in the volatility equation; $\alpha$ -- score coefficient; $\varphi$ -- autoregressive coefficient; $\nu$ -- degrees of freedom; $\pi$ -- zero inflation; $A$, $A_\text{F}$ -- R$^2$ statistic of the ARCH-LM test with lag 10; $\ell$, $\ell_\text{F}$ -- average log-likelihood; $A_\text{F}$, $\ell_\text{F}$ are evaluated on the data from the next day.
\end{flushleft}
\end{table}

\begin{table}
\caption{\textbf{The MSFT stock} -- The median estimated parameters, the ARCH-LM statistic, and the fitted log-likelihood from daily models based on the Student's t-distribution, estimated using various R packages.}
\label{tab:contMSFT}
\centering
\begin{tabular}{lrrrrrrrr}
\toprule
& \multicolumn{4}{c}{1 Second Frequency} & \multicolumn{4}{c}{1 Minute Frequency} \\ \cmidrule(l{3pt}r{3pt}){2-5} \cmidrule(l{3pt}r{3pt}){6-9}
& \verb"rugarch" & \verb"fGarch" & \verb"GAS" & \verb"gasmodel" & \verb"rugarch" & \verb"fGarch" & \verb"GAS" & \verb"gasmodel" \\ 
\midrule
$\mu$ & 0.014 & 0.012 & 0.012 & 0.012 & 0.209 & 0.210 & 0.185 & 0.183 \\ 
$\omega$ & 0.178 & 0.540 & 0.010 & 0.000 & 6.751 & 6.802 & 0.010 & 0.000 \\ 
$\alpha$ & 0.070 & 0.118 & 0.110 & 0.102 & 0.068 & 0.072 & 0.188 & 0.185 \\ 
$\varphi$ & 0.916 & 0.860 & 0.995 & 0.996 & 0.909 & 0.908 & 0.999 & 0.999 \\ \vspace{2mm}
$\nu$ & 4.334 & 4.298 & 4.268 & 4.299 & 7.080 & 6.988 & 8.176 & 8.174 \\ 
$A$ & 0.008 & 0.005 & 0.082 & 0.082 & 0.020 & 0.020 & 0.024 & 0.022 \\ 
$\ell$ & -2.673 & -2.672 & -2.671 & -2.671 & -4.256 & -4.256 & -4.257 & -4.257 \\ 
\bottomrule
\end{tabular}
\begin{flushleft}
\scriptsize
\textit{Notes:}
$\mu$ -- expected price difference; $\omega$ -- intercept in the volatility equation; $\alpha$ -- variance/score coefficient; $\varphi$ -- autoregressive coefficient; $\nu$ -- degrees of freedom; $A$ -- R$^2$ statistic of the ARCH-LM test with lag 10; $\ell$ -- average log-likelihood.
\end{flushleft}
\end{table}

\begin{table}
\caption{\textbf{The MSFT stock} -- The median estimated parameters, the ARCH-LM statistic, and the fitted log-likelihood from daily models based on various integer distributions.}
\label{tab:integerMSFT}
\centering
\begin{tabular}{lrrrrrrrr}
\toprule
& \multicolumn{4}{c}{1 Second Frequency} & \multicolumn{4}{c}{1 Minute Frequency} \\ \cmidrule(l{3pt}r{3pt}){2-5} \cmidrule(l{3pt}r{3pt}){6-9}
& Normal & t & Skellam & Z-I Sk.& Normal & t & Skellam & Z-I Sk. \\ 
\midrule
$\theta$ & -0.396 & -0.330 & -0.361 & -0.384 & -0.026 & -0.027 & -0.013 & -0.013 \\
$\omega$ & 3.393 & 2.109 & 2.559 & 2.640 & 5.999 & 5.816 & 4.320 & 4.320 \\
$\alpha$ & 0.086 & 0.254 & 0.060 & 0.060 & -0.072 & 0.038 & 0.028 & 0.028 \\
$\varphi$ & 0.972 & 0.927 & 0.985 & 0.985 & 0.984 & 0.969 & 0.999 & 0.999 \\
$\nu$ &  & 4.605 &  &  &  & 10.700 &  &  \\ \vspace{2mm}
$\pi$ &  &  &  & 0.080 &  &  &  & 0.001 \\
$A$ & 0.005 & 0.013 & 0.004 & 0.003 & 0.023 & 0.023 & 0.036 & 0.037 \\ \vspace{2mm}
$\ell$ & -2.774 & -2.618 & -2.672 & -2.642 & -4.224 & -4.228 & -4.352 & -4.352 \\
$A_\text{F}$ & x & 0.014 & 0.006 & 0.004 & 0.048 & 0.041 & x & x \\
$\ell_\text{F}$ & x & -2.628 & -2.837 & -2.813 & -6.529 & -4.710 & x & x \\
\bottomrule
\end{tabular}
\begin{flushleft}
\scriptsize
\textit{Notes:}
$\theta$ -- moving-average coefficient; $\omega$ -- intercept in the volatility equation; $\alpha$ -- score coefficient; $\varphi$ -- autoregressive coefficient; $\nu$ -- degrees of freedom; $\pi$ -- zero inflation; $A$, $A_\text{F}$ -- R$^2$ statistic of the ARCH-LM test with lag 10; $\ell$, $\ell_\text{F}$ -- average log-likelihood; $A_\text{F}$, $\ell_\text{F}$ are evaluated on the data from the next day.
\end{flushleft}
\end{table}

\end{document}